# Energetically Favored One-Dimensional Moiré Superstructure in the Pseudo-Square Lattice GdTe$_3$

Jieun Yeon[1,†], Kihyun Lee[1,†], Myeongjin Jang[1], Tae Keun Yun[1], Jongho Park[2,3], Changyoung Kim[2,3], and Kwanpyo Kim[1,*]

[1]Department of Physics, Yonsei University, Seoul 03722, Korea

[2]Department of Physics & Astronomy, Seoul National University, Seoul, 08826, Korea

[3]Center for Correlated Electron Systems, Institute for Basic Science, Seoul 08826, Korea

*Corresponding e-mail: kpkim@yonsei.ac.kr

†Equal Contributions

**Abstract**

**Moiré engineering in layered crystals has recently gained considerable attention due to the discovery of various structural and physical phenomena, including interfacial reconstruction, superconductivity, magnetism, and distinctive optoelectronic properties. Nevertheless, most explored moiré systems have been limited to hexagonal lattices, thereby constraining a comprehensive understanding and technological application of moiré phenomena in general layered crystals. Here, we investigate $GdTe_3$, a pseudo-tetragonal layered crystal, as a platform to explore unconventional moiré phenomena. $GdTe_3$ exhibits a slight in-plane distortion correlated with the direction of charge density wave formation. Through vertical stacking of layers with different distortions—induced via a controlled strain/release process—we realize energetically favorable one-dimensional (1D) moiré superstructures. Using transmission electron microscopy (TEM), including high-resolution scanning TEM imaging, dark-field TEM imaging, and sample tilting experiments, we systematically examine stacking variations across the 1D moiré structure. Additionally, electron energy loss spectroscopy reveals modulations in electronic properties associated with the 1D moiré structure. Our findings expand the scope of moiré systems beyond conventional hexagonal twistronics, enabling exploration of moiré phenomena in low-symmetry van der Waals crystals.**

**Introduction**

Moiré patterns, formed by the superposition of two or more lattices, have attracted intense research interest due to their ability to induce significant modifications in electronic and optical properties.[1-7] In particular, moiré structures observed in twisted hexagonal lattices, such as graphene, h-BN, and transition metal dichalcogenides, have demonstrated remarkable emergent phenomena including superconductivity,[1,4] tunable magnetism,[8-10] and excitonic effects.[11,12] Researchers have typically engineered moiré pattern periodicity and shape through adjustment in unit cell size mismatch,[7] twist angle,[1,13-15] or interlayer sliding.[16] Although the periodicity is largely controllable through the twist angle, the inherent symmetry of the underlying unit cells primarily dictates the overall moiré geometry. Compared to hexagonal-lattice-based systems, stable moiré superlattices in low-symmetry, nonhexagonal crystals remain relatively unexplored, limiting our comprehensive understanding and technological utilization of nonhexagonal moiré phenomena.

Recent efforts have expanded moiré research beyond hexagonal lattices, focusing on structures with alternative symmetries. For instance, twisted moiré systems based on the square lattice of perovskites and the rectangular architecture of black phosphorus have been investigated.[17-22] In these studies, the moiré superlattices reflect the underlying unit-cell symmetry, such as square moiré patterns emerging in twisted perovskites. However, controlling the moiré periodicity, especially at low twist angles with large periodicity, remains challenging due to energetically unfavored twisted structures. Consequently, realizing stable, energetically favorable moiré superstructures using nonconventional layered materials is an important and open question.

The rare-earth tritelluride family, $RTe_3$ (R = rare earth element), presents unique structural and electronic characteristics, notably charge density wave (CDW) phenomena accompanied by slight in-plane distortions.[23-31] Specifically, $GdTe_3$ is an emerging material that exhibits high charge carrier mobility[32] and exotic collective excitations.[24,25] The crystal structure of GdTe$_3$ is intimately linked to its CDW states, with a slight in-plane distortion (c ≈ 1.002 a) locked in by the orientation of the CDW along the c* axis. From the top view, the structure appears nearly square, and such pseudo-square layered structures may provide an exceptional platform for exploring unconventional moiré phenomena.

Here, we investigate GdTe$_3$ as a promising candidate to explore unique moiré phenomena arising from lattice distortion. GdTe$_3$ belongs to the orthorhombic space group *Cmcm* and consists of GdTe slabs sandwiched between two Te nets, as shown in **Figure 1a**. The in-plane lattice parameters differ by only ~0.2%, and the lattice is often described as pseudo-tetragonal. A top view of the structure shows that the in-plane distortion is very small, resulting in a nearly square-like appearance, which we refer to as pseudo-square lattice. GdTe$_3$'s intrinsic in-plane distortion, driven by CDW, allows the generation of energetically favorable one-dimensional (1D) moiré patterns upon vertical stacking of layers with slightly different distortions, achieved through a controlled strained-and-released process. Using transmission electron microscopy (TEM), including high-resolution scanning TEM imaging, dark-field TEM (DF-TEM) imaging, and tilting experiments, we systemically analyze the stacking variations within the observed 1D moiré superlattice. Electron energy loss spectroscopy (EELS) analysis further reveals that the electronic properties are modulated by the 1D moiré structure. Our findings expand the realm of moiré systems beyond conventional hexagonal-lattice twistronics, paving the way for innovative directions in low-symmetry van der Waals (vdW) crystals.

**Results**

We prepared two types of TEM samples by mechanical exfoliation/dry-transfer and characterized them by using various TEM techniques. The pristine GdTe$_3$ samples, prepared without the intentional straining process, exhibited selected-area electron diffraction (SAED) patterns consistent with previous studies,[33,34] revealing the CDW satellite peaks (marked by blue arrows), as shown in **Figure 1b**. A unidirectional CDW forms at $q_c \cong 2/7\mathbf{c}^*$, which can be used to distinguish the **a** and **c** axes in the crystal.[27,35,36] In contrast, the strained/released GdTe$_3$ samples—produced through stretching/shearing on a flexible substrate followed by relaxation (**Figures 1d** and **S1**)—display distinct features in both real and reciprocal space. In SAED, an additional set of CDW peaks was observed, yielding two orthogonal CDW satellite peaks, as indicated by the red and blue arrows in **Figure 1c**. Furthermore, major regions of the strained/released samples exhibited periodic fringe patterns over large area as shown in **Figure 1e**. The local periodicity *d* of the observed fringe was well defined, typically on the order of tens of nanometers.

To understand the formation of 1D fringes and bidirectional CDW ordering in strained/released GdTe$_3$, we conducted a detailed analysis of the SAED, as shown in **Figure 1f-1i**. The enlarged Bragg peaks reveal slight splitting along the horizontal direction in SAED (**Figure 1f and 1g**), which is perpendicular to the fringe lines observed in real space. The splitting of the Bragg peaks was found to be proportional to the vertical distance from the central peak in SAED, but not correlated with the horizontal position, as shown in **S2**. In contrast, the higher-order diffraction peaks along the horizontal direction, such as $(60\bar{6})$, did not show noticeable peak splitting (**Figure 1h**). Moreover, the zoomed-in analysis revealed that the two orthogonal CDW satellite peaks were associated with different reciprocal lattice positions, as illustrated in the schematic on the left of **Figure 1i**.

From the observed diffraction peak splitting, we hypothesized that the strained/released GdTe$_3$ contains two different crystal variants. These two variants with slight distortion, reconstructed based on diffraction information, are shown on the right side of **Figure 1i**. In real-space from top-view, the two variants share the same lattice orientation along the vertical direction, while the horizontal axes exhibit a slight deviation. The observed small splitting angle, $2\alpha$, indicates that the two distorted variants remain very close to a square lattice (**Figure 1i**). To aid visualization, the distortions from the square are not to scale and are intentionally exaggerated in **Figure 1i** and in several subsequent figures. We note that the two variants are related by a $(90 \pm \alpha)$ ° rotation or mirroring. For example, if we assign the unit vectors of the structure as $\mathbf{r}_1$, $\mathbf{r}_2$ and $\mathbf{r'}_1$, $\mathbf{r'}_2$, these vectors are related by relative rotation of $(90 - \alpha)°$, as shown in **Figure 1i**. We hypothesize that the strain-and-release process induces a nonuniform strain field and interlayer sliding at various locations along the thickness direction, resulting in a change in the distortion direction in parts of the sample.

**Figures 2a, b** shows the representative TEM images of 1D moiré patterns observed in strained/released samples, with a periodicity of approximately 40 nm. Contrary to HAADF-STEM imaging, the sharp lines in moiré can display dark contrast in bright-field TEM (BF-TEM) imaging under certain defocus conditions. At this point, it is natural to interpret the observed 1D fringes as moiré formation by the vertical stacking of two slightly different variants. The moiré periodicity $d$ is expected to follow the relationship with the angular splitting $\alpha$ of the Bragg peaks in SAED as follows

$$d \text{ [nm]} = 1/(6.54 \tan \alpha).$$

We analyzed numerous samples and compiled the observed moiré periodicity $d$ and angular splitting $\alpha$, as shown in **Figure 2c**. The experimental data were consistent with the expected $d$–$\alpha$ relationship, confirming that the observed patterns can be explained by vertical stacking of two variants. We note that the experimental measurement of small $\alpha$ below 0.2 ° is challenging due to the significant overlap of the two peaks, as indicated in **Figure 2c**. In the case near $\alpha \approx$ 0.2 °, $\alpha$ was estimated by deconvoluting the broadened peaks using Gaussian fitting, with the results shown as blue squares.

To quantify the dominant structural configurations, we constructed a histogram of the cumulative area of all detected repeating moiré domains as a function of their corresponding periodicity $d$ (**Figure 2d**). This analysis enables the identification of the most energetically favorable and spatially dominant moiré configurations, which reflect underlying structural distortions. Moiré patterns were most frequently observed within an $\alpha$ range of approximately 0.13°–0.27°, corresponding to moiré periodicities of approximately from 32 to 67 nm. Based on this observation, we estimated the a/c ratio to lie within the range from 0.998 to 0.995, which closely matches typical a/c ratios reported for GdTe$_3$.[30] The fact that moiré periodicities outside this range are rarely observed demonstrates that such configurations are energetically unfavorable, likely due to the excessive structural distortion required to sustain them.

To understand the stacking energetics, we calculated the interfacial potential energy as a function of the relative rotation angle $\theta$ between two overlapping GdTe$_3$ crystals using an analytical model (see Methods for details) as shown in **Figures 2e-f** and **S3**. To simplify the system, we considered two overlapping Te nets with relative rotation, as shown in **Figure 2e**. Each Te net is assumed to be slightly distorted from a square, with corner angles of $(90\pm\alpha)°$. The calculations, with an exemplary case of $\alpha = 0.3°$, reveal that energetically favorable configurations occur at $\theta = 0°$ (trivial case) and near $\theta = 90°$ (red-colored region). A zoomed-in view of the potential near $\theta = 90°$ shows that the favored configurations appear at $\theta = (90\pm\alpha)°$, as illustrated in **Figure 2f**. A previous calculation based on density functional theory[27] suggests that the energy stabilization for $\theta = (90\pm\alpha)°$ is approximately 20 mJ/m$^{-2}$. **Figure 2g** presents a schematic structure corresponding to the $\theta = (90-\alpha)°$ case, where the formation of 1D moiré patterns is apparent. Overlapping lattice positions are marked in yellow, highlighting the emergence of a 1D moiré pattern. We can assign regions with dark contrast (**Figure 2a, b**) to dislocation cores with imperfect crystal alignment, which induce strong electron scattering. In contrast, in HAADF-STEM—where electrons scattered at high angles are collected—the contrast is reversed, as shown in **Figure 1e**. The overlaid structure can be interpreted as a 1D array of dislocations, with Burgers vector **B** aligned parallel to the 1D moiré direction and exhibiting screw dislocation character.[37] The observed Burgers vector **B** = (1/2 ***a*** + 1/2 ***c***) is consistent with a recent study.[27] Dark-field TEM is an excellent technique for investing in the dislocation characters and local stacking variations. **Figure 3a** shows a BF-TEM image of the 1D moiré structure used for DF-TEM analysis. In the BF-TEM image, periodic sharp dark lines are observed, corresponding to structurally aligned regions. Along these lines, screw dislocations with Burgers vectors **B** (red arrows) are aligned parallel to the moiré fringes.

DF-TEM contrast strongly depends on the selected diffraction spots (**Figure 3b**). **Figure 3c–e** compares the experimental and simulated dislocation contrasts in the (202), (200), and (20$\bar{2}$) DF-TEM images, acquired under strong two-beam conditions. The simulated results show excellent agreement with the experimental observations. The (200) DF image exhibits dislocations as bright lines against a dark background (**Figure 3c**). In the (202) DF image, bright lines appear periodically at positions corresponding to half the period of the original 1D moiré pattern (**Figure 3d**). This behavior is consistent with the established interpretation of dislocation contrast, where ***g*** · **B** = 2 leads to a symmetric double-line feature.[38] In this case, the displacement field around the dislocation crosses a phase difference of ± π twice, producing the appearance of two adjacent contrast lines. Notably, the applicability of the ***g*** · **B** = 0 invisibility criterion is demonstrated in the (20$\bar{2}$) DF image, where the corresponding dislocation becomes invisible (**Figure 3e**).

High-angle annular dark-field (HAADF)-STEM imaging was utilized to investigate the local stacking configurations across the 1D moiré at atomic resolution. **Figure 4a** presents a structural model of the 1D moiré pattern across a single moiré period, in which a relative displacement corresponding to a Burgers vector **B**—equivalent to half a unit cell—occurs. In this model, the top and bottom layers undergo lateral displacements in opposite directions, parallel to the 1D moiré lines. An experimentally obtained, high-resolution HAADF-STEM image is shown in **Figure 4b**. Zoomed-in STEM images from different regions display distinct contrast patterns, as shown in **Figures 4d–4g**. A reference region away from 1D moiré lines shows the atomic contrast expected from the bulk crystal (**Figure 4d**). In contrast, near 1D moiré lines—which appear brighter in the HAADF-STEM image—an additional signal originating from displaced layers (marked by blue layers) is observed, as shown in **Figures 4e–g**. Here, the layers represented by blue have a weaker signal compared to pink layers due to different thickness values in samples. The simulated HAADF-STEM image, based on the vertical overlap of two structures is in a good agreement with experimental data, as shown in **Figure 4c**.

The side plots in **Figures 4d–g** display line profiles extracted from the cropped experimental images, revealing that the background signal at the moiré fringe is not randomly distributed but instead appears as a shoulder peak consistently oriented in the direction with linear displacement across all domains. To measure the relative shift between the top and bottom sample segments, we extracted and tracked the relative shift of the fitted low-amplitude peaks with respect to the corresponding blue signals. The relative shift is in a good agreement from the simple model adapting the overlap between two distortion variants (**Figure 4h**). We note that the experimental identification of local stacking was performed only near the central region due to the significant overlap between two layers in areas farther from the central regions.

Tilting measurements provide out-of-plane structural information that cannot be accessed from a top-view projection alone.[3,39] **S4** presents the HAADF-STEM images of the moiré region at different tilting conditions. When precisely aligned along the (010) zone axis, the intensity distribution exhibits mirror symmetry across the fringe. Upon tilting the sample along the indicated axis, the brightest moiré fringe shift laterally, in the direction perpendicular to 1D moiré lines (**S4**). This lateral shifts of the misaligned region arises from the breaking of translational symmetry induced by the stacking of the two distorted variants. Furthermore, the direction (left or right) of the moiré shift depends on the stacking configuration between the top and bottom structures. Therefore, TEM analysis with tilting samples provides critical vertical structural information that is not accessible through top-view imaging along the [010] zone axis.

EELS analysis provides information regarding the local electronic properties of the samples.[40] We measured the EEL spectrum across fringe patterns to investigate whether the stacking configuration influences the electrical properties. To enhance the signal–to–noise ratio of the spectrum, the measured data were cropped into segments containing two periods each (red lines) and averaged as explained in **S5**. The zero loss peak and plasmon signal show low intensity at the moiré line, which is consistent with what we expect from the HAADF-STEM contrast[41] as shown in **S5**. **Figure 5b** shows the core-loss region of Gd, characterized by a prominent Fano-shaped giant resonance peak (black arrow) and a weak pre-edge feature (red arrow). The weak pre-edge feature at $\approx 140\,\text{eV}$ arises from bound $4d \rightarrow 4f$ multiplet transitions.[42-44] The main peak, which exhibits the highest amplitude, originates from the $4d \rightarrow 4f$ electronic transition and is in good agreement with previously reported experimental results.[43,45,46] The presence of the Fano line shape indicates interference between discrete and continuum states, which is a hallmark of resonant excitation in rare-earth elements. From our analysis, no noticeable change in the Gd core-loss signal was observed across the moiré structure.

The Te core-loss signal exhibits location-dependent behavior. **Figure 5c** presents the averaged core-loss EELS spectrum of Te, with the signals from the bright and dark regions in the STEM image plotted in blue and red, respectively. While the two spectra exhibit nearly identical features across most of the energy range, a clear difference is observed near the onset of the Te core-loss signal. The inset in **Figure 5c** show a magnified view of the onset region, including both the averaged raw data and the linear fitting results, highlighting the slope difference between the bright and dark regions. From the linear fitting, the onset energy of the Te signal was extracted, as shown in **Figure 5d**. The green-shaded areas correspond to the bright regions indicated in **Figure 5a**, and the onset values in these regions exhibit a decreasing trend. The extracted onset energies from the dark and bright regions shows an average difference of approximately 0.3 eV as shown in the inset of **Figure 5d**. Moreover, a normalized difference plot obtained by subtracting the EELS spectrum of the bright region from that of the dark region clearly visualizes the spectral variations between the two (**Figure 5e**). The most prominent difference appears in the energy range of 595–599 eV, where the slope of the difference signal is the steepest, indicating a significant change in spectral intensity.

The onset and shape of the observed core-loss EELS signal are closed linked to the unoccupied electronic density of states near the Fermi level in materials.[40] In GdTe$_3$, the electronic states near the Fermi level are dominated by the Te $p$-orbitals in the Te nets, which are sensitive to the inter-net stacking geometry and any associated slight distortions.[29,47] Previous studies have also indicated that interlayer shifting and cleaving between Te net layers requires relatively low energy.[32,48] Therefore, the observed variation in the Te core-loss signal can be attributed to modulation of electronic properties associated with stacking variation between Te nets in the moiré structure. As the stacking sequence gradually changes across the screw dislocation core, the Te nets experience interlayer sliding. In particular, the bright regions correspond to locations where the interlayer stacking configurations between Te nets are displaced relative to the bulk counterpart.

**Conclusions**

In conclusion, we have demonstrated the formation of energetically favored 1D moiré superstructures in the pseudo-square lattice GdTe$_3$. Using strain-induced stacking changes, we produced two distorted variants, whose vertical stacking leads to periodic 1D-moiré fringes and associated bidirectional CDW formation. Through comprehensive structural analysis—including SAED, DF-TEM, HAADF-STEM imaging, simulations, and tilting experiments—we confirmed the formation of energetically favored 1D moiré patterns and revealed local stacking configurations. Additionally, EELS measurements revealed electronic modulation across the moiré, suggesting that this system holds potential for tuning magnetic and electronic properties via stacking configuration. These findings not only highlight the potential of strain-engineered twistronics in rare-earth tellurides but also establish a framework for controlling quantum properties in nonhexagonal layered vdW systems, thereby enabling exploration of moiré systems with functional diversities based on low-symmetry crystals.

**Methods**

**$GdTe_3$ Crystal Growth.** Stoichiometric single-crystalline $GdTe_3$ flakes were synthesized by the self-flux growth method using a box furnace.[25] High-purity Gd metal (99.9 %) and Te chips (99.999 %) were mixed in a 1:30 molar ratio for the Te-rich self-flux condition. The mixed precursor of $GdTe_3$ was loaded into quartz tubes and sealed at ~$10^{-5}$ Torr by using a turbo pump to prevent oxygen contamination. The maximum heating temperature was set to 900 °C, and the temperature was maintained for 24 h to achieve a homogeneous melt. After melting, the sample was cooled down to 500 °C at a rate of – 2 °C/hour. Melt in an ampule was decanted in the room temperature until it turns solid. The crystal structure was determined using a high-resolution X-ray diffractometer with an airtight holder to prevent degradation caused by oxygen.

**TEM Measurement and Simulations.** TEM samples were prepared using the poly(methyl methacrylate) (PMMA)-based transfer method. Mechanical exfoliation, transfer to a holey $Si_3N_4$ membrane TEM grid, and optical characterization were performed in a $N_2$-filled glovebox.[49] The sample thickness investigated by TEM was typically on the order of a few tens of nanometers. Atomic resolution HAADF-STEM images, SAED patterns, and EELS were acquired using a spherical (Cs)-aberration corrected JEOL ARM-200F operated at 200 kV. TEM images, DF-TEM images and additional SAED patterns were obtained with JEOL-2100PLUS operated at 200 kV. EELS analysis was conducted using a 965 GIF Quantum dual EELS spectrometer (Gatan) at an accelerating voltage of 200 kV. The energy dispersion was 0.1 eV/channel with a convergence semiangle of 23 mrad, and the collection semiangle for EELS was 4 mrad. The full width at half maximum (FWHM) of the zero-loss peak in the measurement was approximately 0.8 eV. STEM image simulations were performed by using the computem software. The simulations were conducted based on multislice calculations for HR-STEM imaging, incorporating experimental parameters such as 200 kV acceleration voltage, a probe convergence angle of 25 mrad, and electron counting noise. SAED patterns and DF-TEM images are simulated using the abTEM[50] simulation package. To simulate the 1D moiré structure, we considered a gradually changing stacking sequence by applying incremental sliding shifts at the Te net interface. For DF-TEM simulations, the intensity profiles of each diffraction peak were calculated and used to generate the corresponding 2D imaging contrasts. A total of ten layers of $GdTe_3$ were modeled, in which the bottom five layers (1L–5L) remained fixed, while the top five layers (6L–10L) were gradually displaced in a direction parallel to the Burgers vector.

**Calculation.** The relation between the moiré periodicity $d$ and angular splitting angle $\alpha$ in electron diffraction was derived based on geometric considerations in diffraction, and the detailed calculation is given in **Supporting Information Note 1** and **Figure S6**. The interfacial potential energies between two $GdTe_3$ structures were calculated using an analytical method [51,52]. A detailed description of the calculation is provided in **Supporting Information Note 2**.

**Acknowledgements**

This work was supported by the National Research Foundation of Korea (NRF) grant funded by the Korea government (MSIT)(RS-2025-00560649) and Yonsei Signature Research Cluster Program of 2024 (2024-22-0004). This work was supported by the Global Learning & Academic research institution for Master's·PhD students, and Postdocs (G-LAMP) Program of the National Research Foundation of Korea grant funded by the Ministry of Education (No. RS-2024-00442483).

Supporting Information

The Supporting Information is available free of charge at https://pubs.acs.org/doi/10.1021/acsnano.5c09994 .

List Item Experimental and analysis details including sample fabrication process, optical image of samples, Bragg peak splitting analysis, interfacial potential calculations, sample tilting experiments, and additional EELS data (PDF)

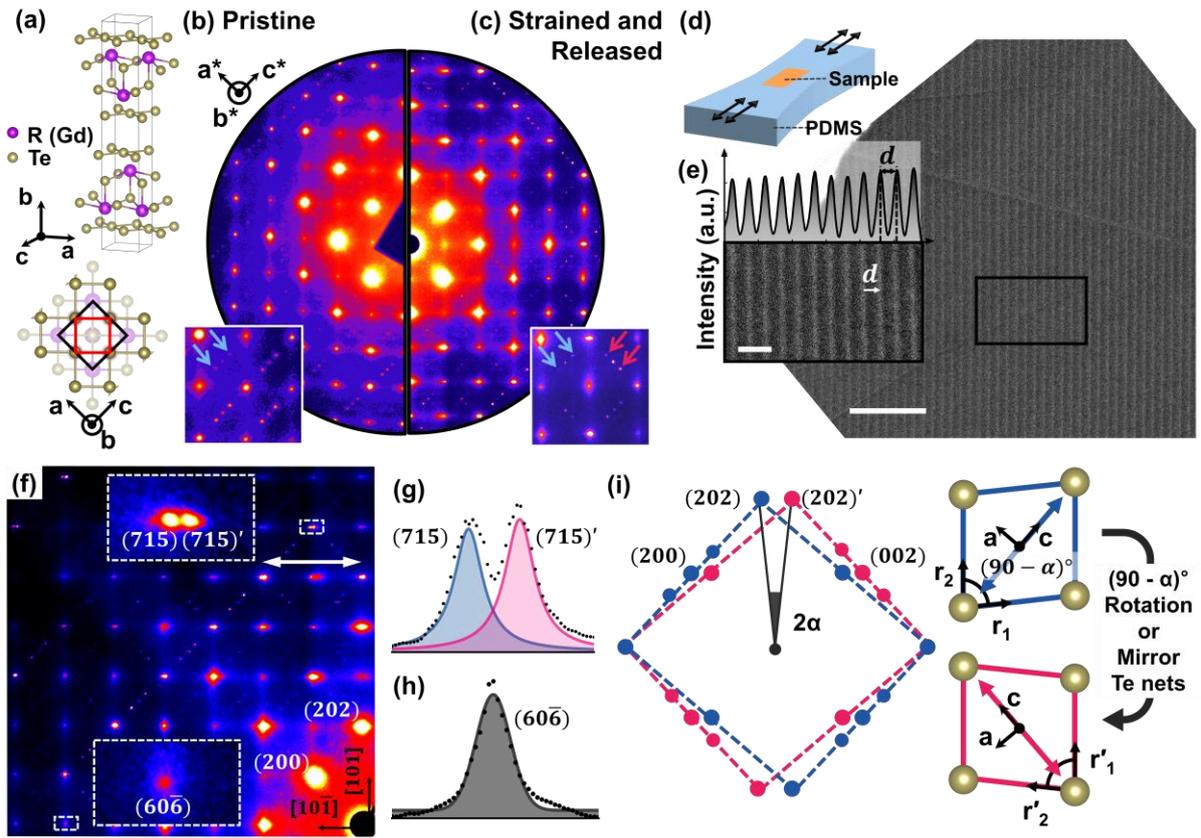

**Figure 1. 1D moiré in strained–and–released GdTe$_3$.** (a) Perspective and top views of the GdTe$_3$ crystal structure. The black square represents the unit cell, and the red square indicates the Te net. (b, c) SAED patterns of pristine and strained/released GdTe$_3$ samples along the [010] zone axis. Insets are the enlarged view of the SAED with marked CDW peaks. (d) Schematic illustration of the PDMS-assisted strained-and-released sample preparation. (e) TEM image showing periodic fringe patterns in the strained-and-released sample. The line intensity profile across the fringe pattern is shown at the top of the enlarged TEM image. In the main and enlarged TEM images, the scale bars correspond to 200 nm and 50 nm, respectively. (f) SAED pattern of the strained/released GdTe$_3$ sample. Magnified views of the (715) and (60$\bar{6}$) Bragg peaks are shown inside the dashed boxes. (g, h) Line intensity profiles corresponding to the (715) and (60$\bar{6}$) peaks. (i) Left: schematic showing overlapping SAED peaks from two rotational variants (blue and read), where a small angular splitting in the (202) peaks is denoted by 2α°. Right: Schematic illustration of two Te nets based on the observed SAED, which are related by a (90 ± α)° rotation or mirror symmetry. The distortion from the square is not to scale and is intentionally exaggerated to aid visualization.

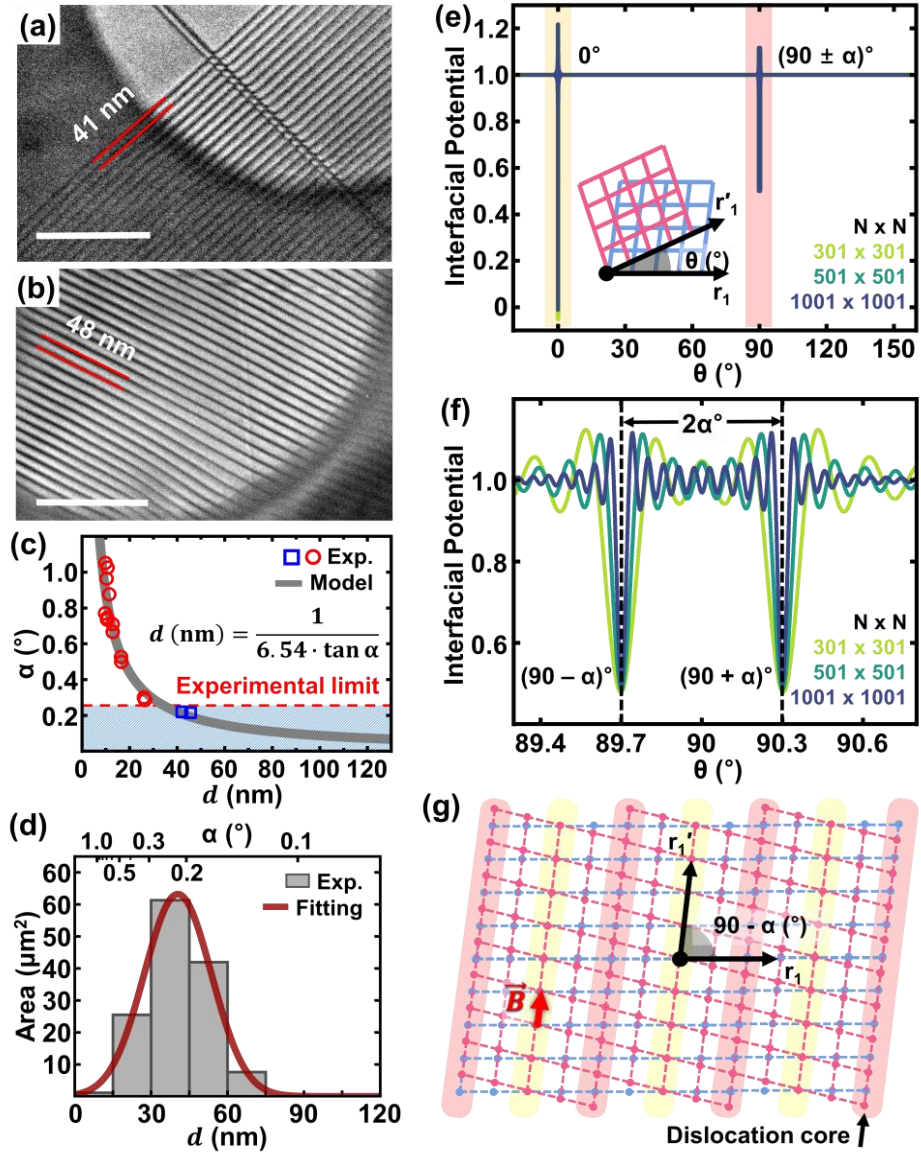

**Figure. 2 Energetically favorable 1D moiré formation.** (a,b) Exemplary TEM images showing 1D moiré. Scale bar: 500 nm. (c) Experimentally measured moiré periodicity $d$ and angular splitting $\alpha$ in electron diffraction. The relation between $d$ and $\alpha$ arising from 1D moiré formation via vertical stacking is shown as a gray line. The experimental limitation in measurement of $\alpha$ is indicated. (d) Histogram of observed moiré periodicity $d$ and angular splitting $\alpha$. (e) Calculated interfacial potential energy as a function of the rotational angle $\theta$ between two Te lattices, where $N$ represents the number of Te unit cells. (f) Magnified plot of the interfacial potential near $\theta = 90°$ in (e). (g) Schematic of the 1D moiré pattern structure, corresponding to the energetically stable angle of $(90 - \alpha)°$ presented in (f). The overlapping lattice positions is marked in yellow, showing the formation of the 1D moiré pattern. Burgers vector **B** is marked with its direction parallel to the 1D moiré.

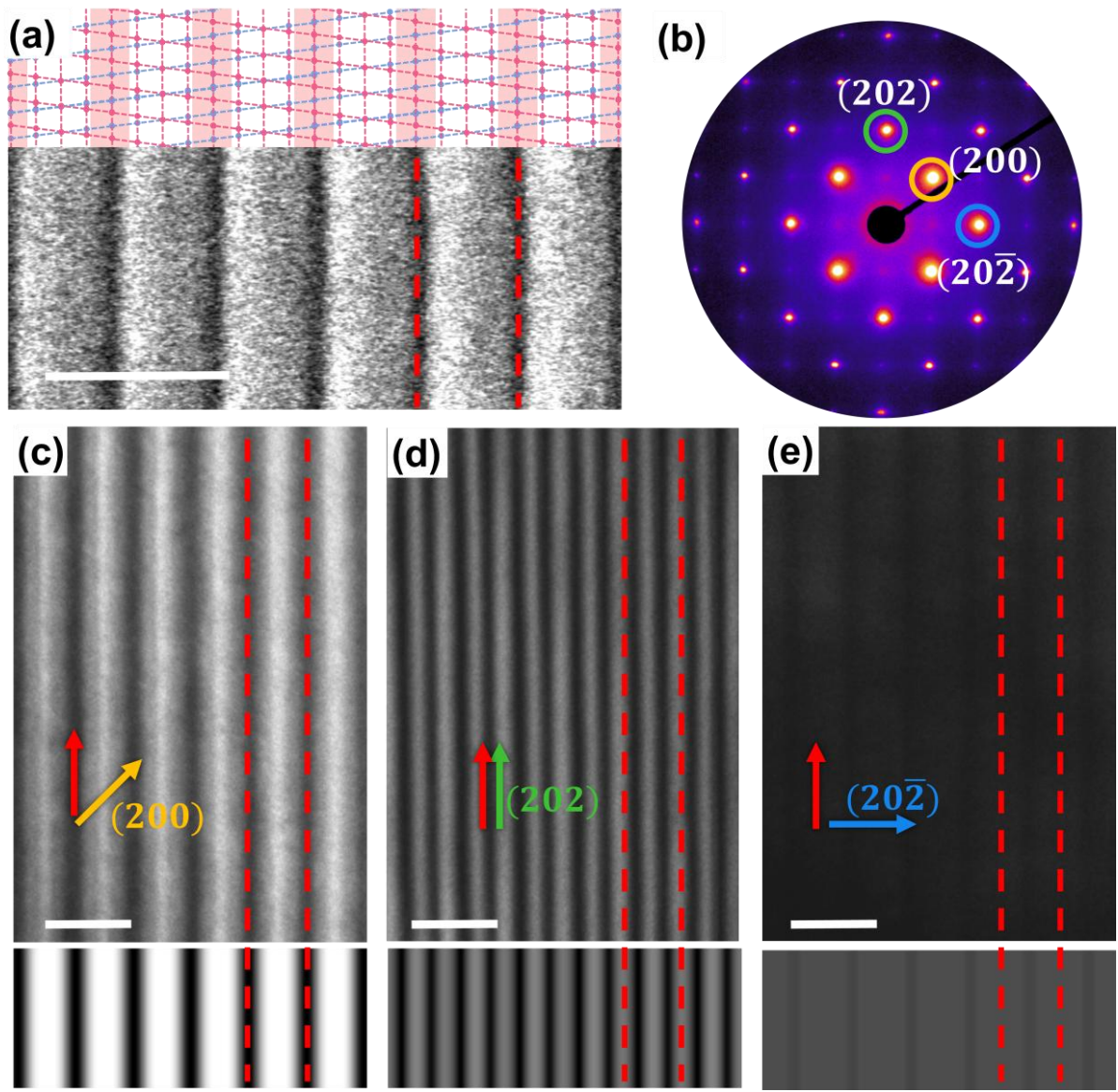

**Figure 3. Dark-field (DF) imaging of 1D moiré structures.** (a) Bright-field (BF) TEM image of the 1D moiré sample used for DF-TEM measurements, along with a schematic illustration of the structure. Scale bar: 50 nm. (b) SAED pattern indicating selected diffraction spots used for DF TEM imaging. Each peak is selected for dark-field imaging in panels (c–e). (c–e) Experimental DF-TEM images (top) obtained using the (200), (202), and ($\bar{2}$02) diffraction spots, respectively, with corresponding simulated DF-TEM images (bottom). Scale bar: 50 nm. TEM images (bottom). Scale bar: 50 nm.

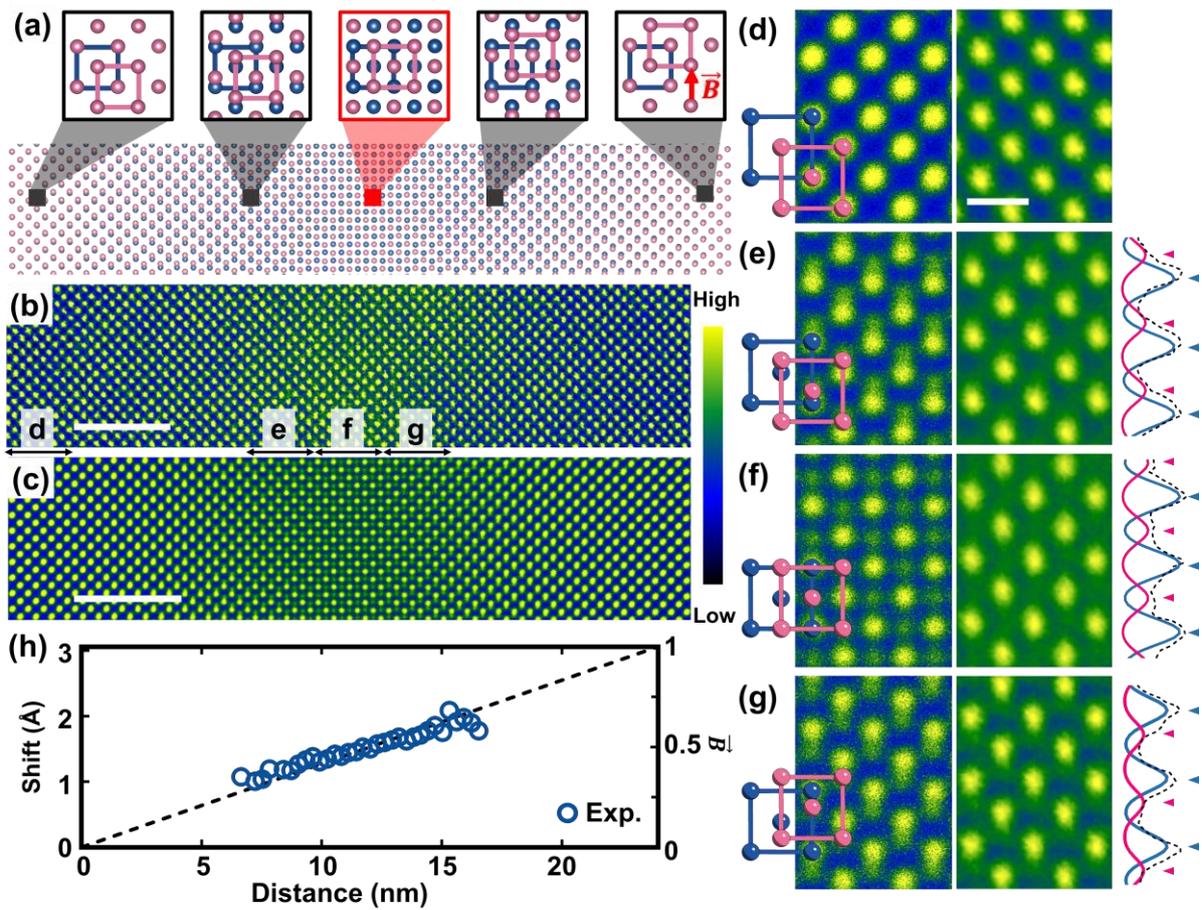

**Figure. 4 High-resolution STEM imaging and analysis of stacking sequence across a 1D moiré.** (a) Structural model illustrating the relative shift between top and bottom GdTe$_3$ across the 1D moiré pattern. Red and blue squares indicate the interfacial Te nets from the top and bottom overlapping layers, respectively. (b) Experimental high-resolution HAADF-STEM image showing one repeating period of the 1D moiré pattern. Scale bar: 2 nm. (c) Simulated HAADF-STEM image corresponding to the same region. Scale bar: 2 nm. (d–g) Enlarged views of simulated (left) and experimental (right) STEM images from selected regions of panels (b) and (c). Scale bar: 3 Å. Overlaid structural models of the two Te nets are shown on the simulated images. Vertical intensity profiles are presented on the right side of the corresponding experimental STEM images. (h) Experimentally measured relative shift between the two overlapping layers across the 1D moiré.

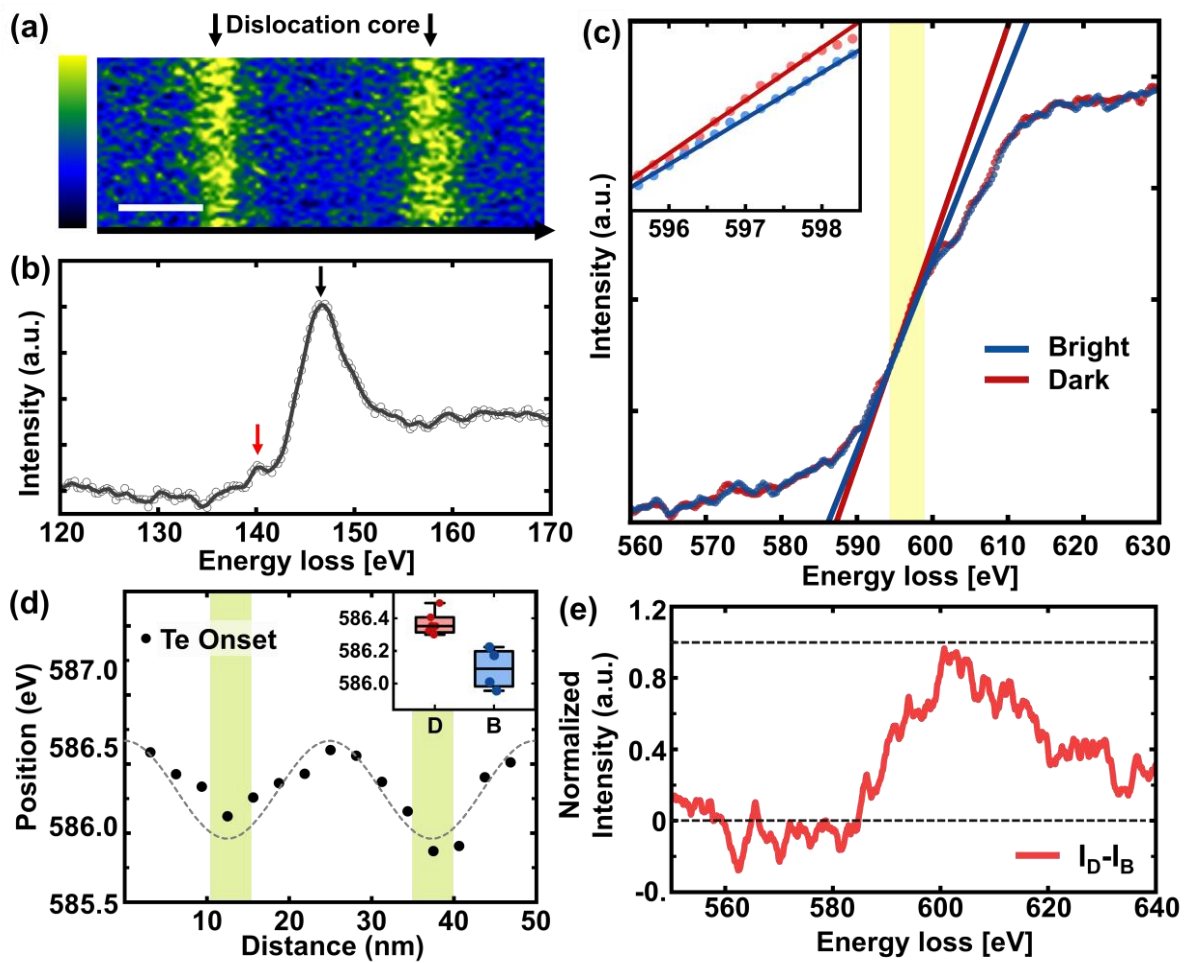

**Figure 5. Electron energy loss spectroscopy (EELS) analysis of the 1D moiré.** (a) HAADF-STEM image of the 1D moiré, showing alternating bright and dark regions. Scale bar: 20 nm. (b) Averaged Gd core-loss EEL spectrum, highlighting the pre-edge peak (red arrow) and the giant resonance peak (black arrow). (c) Averaged Te core-loss spectrum from bright regions (blue) and dark regions (red). Linear fitting was applied at the onset of the $M_4$ edge. The inset shows an enlarged view near the Te edge onset region. (d) Plot of the Te onset positions across the 1D moiré. Green-shaded areas correspond to the bright regions shown in panel (a). The inset shows a box plot comparing the onset values in the dark (D) and bright (B) regions. (e) Normalized difference in Te core-loss intensities between the bright and dark regions.